\documentclass[12pt, a4paper, twoside]{fec}
\title{Numerical relaxation of a 3D MHD Taylor - Woltjer state subject to abrupt expansion}

\author[]{Rupak Mukherjee}
\author[]{Rajaraman Ganesh}
\affil[]{Institute for Plasma Research, HBNI, Bhat, Gandhinagar 382428, INDIA}

\doi

\emailID{ganesh@ipr.res.in}

\PaperNumber{TH/P5-3} 

\begin{document}

\maketitle

\begin{abstract}
{Since the advent of Taylor-Woltjer theory [J B Taylor, PRL, 33, 1139 (1974), L Woltjer, PNAS, 44, 489 (1958)], it has been widely believed that situations with perfectly conducting boundaries and near ideal conditions, the final state of MHD system would be force-free Taylor-Woltjer states defined as $\vec{\nabla} \times \vec{B} = \alpha \vec{B}$ with $\alpha$ as a constant and $\vec{B}$ is the magnetic field defined over a volume $V$. These states are of fundamental importance in fusion plasmas [J B Taylor, RMP 58, 741 (1986)]. More recently, several new MHD models have been proposed - for example Reduced Multi-region relaxed MHD [S R Hudson {\it et al}, Phys. Plasmas, 19, 112502 (2012)] and arbitrary scale relaxation model to Taylor-Woltjer state [H Qin {\it et al}, PRL, 109, 235001 (2012)] to mention a few.

In the present work, we use a 3D compressible MHD solver in cartesian geometry which can handle conducting or periodic as well has mixed boundary conditions to investigate numerically the arbitrary scale relaxation model proposed by Qin et al [H Qin {\it et al}, PRL, 109, 235001 (2012)]. For this purpose, we consider two volumes $V_{init}$ and $V_{final}$. We load the 3D MHD solver in the limit of zero compressibility with a Taylor-Woltjer state $B_{init}(x,y,z,t=0)$ and let it again a numerical evolve with conducting boundaries at $V_{init}$ to make sure that we have obtained a numerically steady Taylor - Woltjer state for volume $V_{init}$. Followed by this procedure, we ``suddenly'' relax the boundaries to a new volume $V_{final}$, such that $V_{init}$ $<$ $V_{final}$ and evaluate whether or not the system attains quasi-steady state. Details of the numerical method used, the protocol followed, the expansion technique and the novelty of this numerical experiment and details of our results have been presented in this paper.}
\end{abstract}

\section{Introduction}

The steady state or late time state of a system of ideal gas can be easily predicted by the extremising the free energy of the system defined as $F = U - TS$, where, $U$ is the internal energy of the system and $T$ represents the absolute temperature and $S$ the entropy of the gas. Extremisation methods of the general steady states arising out of such arguments are powerful tools to understand the underlying physics issues in a global way. One such principle was derived by L. Woltjer [\ref{Woltjer:58}] in 1954 wherein for a near ideal plasma the free energy constructed out of magnetic helicity ($H_m = \int \vec{A} \cdot \vec{B} dV$) and magnetic energy ($E_m = \int B^2 dV$ and $\vec{B} \cdot \hat{n} = 0$) is extremised. This particularly yielded $\vec{\nabla} \times \vec{B} = \alpha(x) \vec{B}$ subject to $\vec{B} \cdot \vec{\nabla} \alpha = 0$ on the ``walls". $\alpha = \alpha(x)$ represents the rigidity of large number of volume ($H_m = \int\limits_V \vec{A} \cdot \vec{B} dV$ where, $V$ is the nested volume).\\

J. B. Taylor [\ref{Taylor:74}] considered a particular limit where, a weak dissipation would break all the local helicity constants except the one considered over the vessel with conducting surfaces, resulting in $\alpha(x) = \alpha_0 = constant$ and thereby, $\vec{\nabla} \times \vec{B} = \alpha_0 \vec{B}$.\\


A whole new set of experimental device emerged out of this study called Reversed - Field - Pinch (RFP) devices for fusion plasma studies. In cartesian geometry such force-free state exists and the states are called Beltrami-type states. But the key point of the plasma relaxation procedure was the fact that the relaxation takes place due to addition of small viscosity which is known to act only at the small scales. Thus it was known that large scales do not participate in the plasma relaxation until H. Qin {\it et al} [\ref{Qin:12}] who has given arbitrary scale relaxation model. Many new models of MHD has also been proposed - for example S. R. Hudson {\it et al} [\ref{Hudson:12}].\\

In the late 1990's Mahajan and Yoshida [\ref{Mahajan:98}] put forth a relaxation model which included the plasma flow $(\vec{u})$. Thus they defined a generalised helicity $H_G = \int (\vec{u} + \vec{A}) \cdot (\vec{\omega} + \vec{B}) dV$ and used energy $W = \int (u^2 + B^2) dV$. The final state resulted in a ``Double Beltrami" flow [\ref{Mahajan:98}]. Yet another model of ``minimum energy dissipation rate" subject to constant megnetic helicity ($H_m$) resulted in ``Triple Beltrami" structure [\ref{Dasgupta:98}].\\

In this paper we numerically evolve a three dimensional MHD plasma from a Beltrami class of solution in a region bounded by conducting walls and ``suddenly" allow to expand the plasma and fill the new volume. The expansion mediates via reconnection of magnetic field lines thereby flow of kinetic and magnetic energy between scales. We measure the scales involved during this relaxation of the plasma and from our numerical tools attempt to identify a model of plasma relaxation. We use conducting boundary to ensure the magnetic helicity is uniquely defined (i.e. $\vec{B} \cdot \hat{n} = 0 = \vec{u} \cdot \hat{n}$ where $\hat{n}$ is the unit normal) and the parameters of the numerical runs are so chosen that the magnetic helicity decays very slowly compared to the total energy (kinetic + magnetic) of the system thereby providing a Taylor like condition to work with. The parameters are selected such that the ``back-reaction" of the magnetic field on the flow ``$\vec{u}$" is controlled by $M_A$ - ``Alfven Mach" number. We consider ABC flow with wave number $k$, for both velocity and magnetic field with velocity scale factors determined by $M_A$. It is important to note that ABC flows are Beltrami class of solution\\

We find that,
\begin{enumerate}
\item For the chosen parameter magnetic helicity is almost conserved as compared to total energy, while total helicity $H_G = \int (\vec{u} + \vec{A}) \cdot (\vec{\omega} + \vec{B}) dV$ decays at a comparable rate to energy with time. Thus inspite of the flow, the parameters chosen are such that magnetic helicity ($H_m$) remains constant while $H_G$ decays allowing a Taylor-like situation.
\item For wavenumber $k = 1$, for all values of the parameters studies, power in magnetic energy spectra is seen to increase while the power in kinetic energy spectra decreases with time. 
\item For $k = 4,8$, the behavior is opposite for magnetic energy spectra.
\item As plasma expands from time $t=0$ with $k = 1$ from a small volume to fill up the simulation volume, in general while the whole spectra is seen to contribute, mode numbers $k \sim 10$ are seen to participate in the relaxation process more dominantly.
\item For $k = 4,8$, the relaxation process is dominantly controlled by $k \textgreater 10$ modes.
\end{enumerate}

\section{Governing Equations}

The following equations govern the dynamics of an incompressible MagnetoHydroDynamic plasma.
\begin{eqnarray*}
\label{Momentum} && \frac{\partial \vec{u}}{\partial t} + \vec{u} \cdot \vec{\nabla} \vec{u} - \vec{B} \cdot \vec{\nabla} \vec{B}  = -\vec{\nabla} p + \mu \nabla^2 \vec{u}\\
\label{Induction}&& \frac{\partial \vec{B}}{\partial t} + \vec{u} \cdot \vec{\nabla} \vec{B} - \vec{B} \cdot \vec{\nabla} \vec{u} = \eta \nabla^2 \vec{B}
\end{eqnarray*}
where $\vec{u}$, $\vec{B}$, $p$ represent the velocity, magnetic and magnetic pressure fields, $\mu$ and $\eta$ represent kinematic viscosity and magnetic diffusivity.
The dimensionless numbers are defined as: $Re$ $=$ $\frac{U0 L}{\mu}$, $Rm$ $=$ $\frac{U0 L}{\eta}$, $Pm$ $=$ $\frac{\mu}{\eta}$. The maximum fluid speed is $U_0$ and the Alfven speed is $V_A = \frac{U_0}{M_A}$, where $M_A$ is the Alfven Mach number.\\

We time evolve the above set of equations in three dimensions within a rectangular volume with perfectly conducting surface such that the tangential components of the electric field vanish. This implies that the normal component of the magnetic field vanishes on the surface (i.e. $\hat{n} \cdot \vec{B} = 0$). Further we choose the initial magnetic field profile a Beltrami type in a cartesian geometry which satisfies the force free condition (viz. $\vec{\nabla} \times \vec{B} = \alpha \vec{B}$). In this problem we have taken Arnold - Beltrami - Childress (ABC) flow ( $\alpha = 1$ for ABC flow) given by:
$$B_x = A \sin (k z) + C \cos (k y); ~~ B_y = B \sin ⁡(k x) + A \cos ⁡(k z); ~~ B_z = C \sin (k y) + B \cos (k x)$$
and choose the velocity profile identical to the magnetic field profile.

\section{Details of BMHD3D}

To study the dynamics of the field variables evolved from any typical initial condition a new code ``Three Dimensional Bounded Magneto-Hydro-Dynamics" (BMHD3D) has been developed in house at Institute for Plasma Research (IPR). The basic subroutines are extensively benchmarked with the known results. In order to evaluate the spatial derivatives we use pseudo-spectral method using discrete sine and cosine fourier transform from FFTW library \ref{FFTW}. For time marching Adams - Bashforth algorithm has been used. Depending on the performance of the code, the grid resolution is chosen as $N_x = N_y = N_z = 64$. The system length is set to $L = 2 \pi$ in all the three directions. Initially the plasma is contained in a conducting boundary between $L/3$ to $2L/3$. The velocity and magnetic fields are ABC type to start with satisfying the force-free criteria. Now the plasma is allowed to move in between the conducting walls at $0$ and $L$. We record all the field variables after every $100$ time intervals from the beginning of the simulation till $5 \times 10^4$ time iteration.\\

Depending on the grid resolution we use, the Raynold's numbers are optimised at $Re = 10^3$ and $Rm = 5 \times 10^3$ such that $Pm = 5$. The Alfven Mach number ($M_A$) is kept $\textgreater 1$ for all cases, such that at the beginning of the simulation the kinetic energy is larger than the magnetic energy. To maintain CFL condition we use $U_0 = 0.1$ and $A = B = C = 0.1$ and the width of time-stepping ($\delta t$) = $10^{-2}$. We keep $k = 1$ for all the studies such that the fluctuations in the field variables remain well-resolved within the grid resolution we have chosen.\\

Apart from the field variables the code evaluates vorticity ($\vec{\omega} = \vec{\nabla} \times \vec{u}$), current ($\vec{J} = \vec{\nabla} \times \vec{B}$), casimirs $\left( C_p = \int\limits_V \vec{\omega}^p dV \right)$, Fluid Helicity $\left( H_f = \int\limits_V \vec{u} \cdot \vec{\omega} dV \right)$, Magnetic Helicity $\left( H_m = \int\limits_V \vec{A} \cdot \vec{B} dV \right)$ and the cross Helicities $\left( H_{uB} =  \int\limits_V \vec{u} \cdot \vec{B} dV \text{and} ~ H_{A\omega} = \int\limits_V \vec{A} \cdot \vec{\omega} dV \right)$ at every time iteration. We also explicitly monitor the $\vec{\nabla} \cdot \vec{B} = 0$ condition and found it to be ${\cal{O}}(10^{-14})$ throughout the double-precesion simulation.

\section{Numerical Results}

\begin{figure}
\includegraphics[scale=0.65]{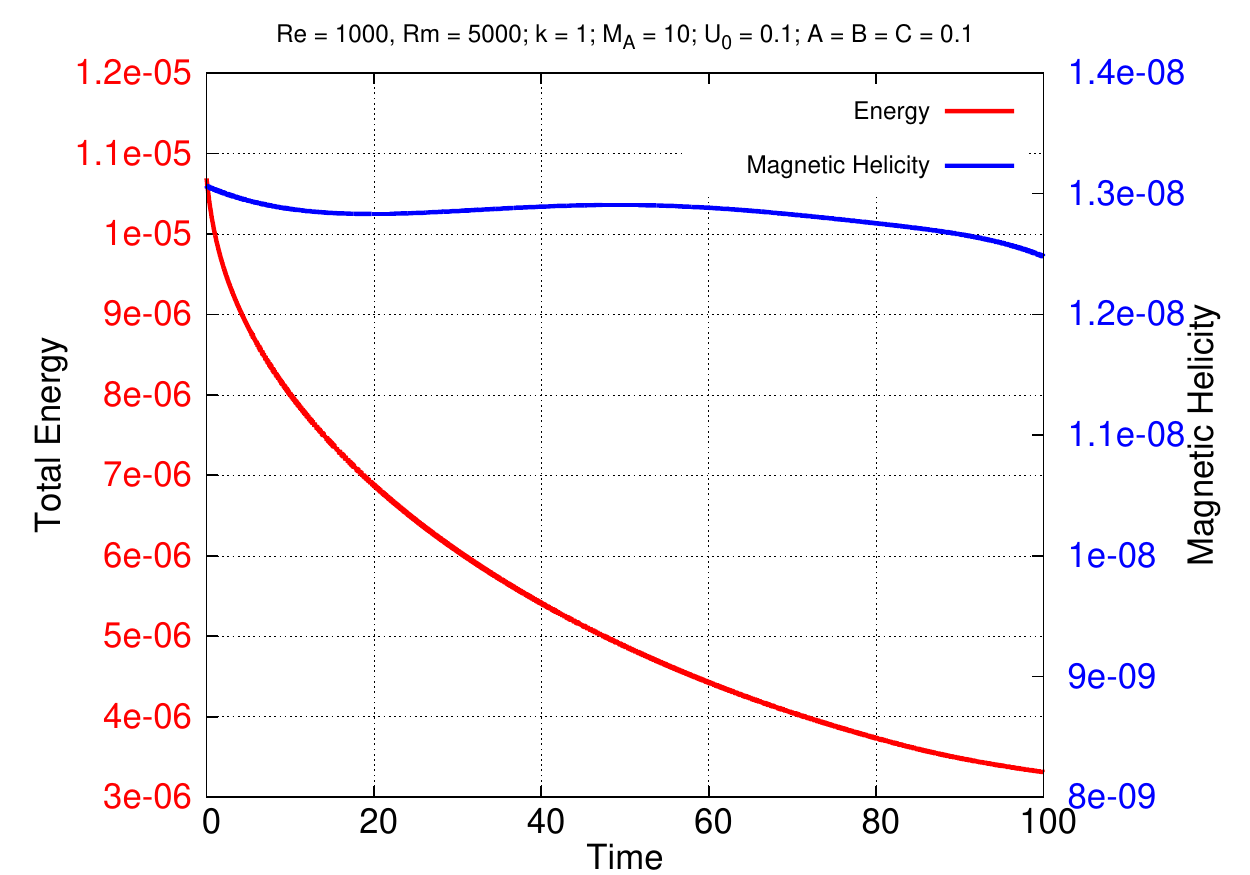}
\includegraphics[scale=0.65]{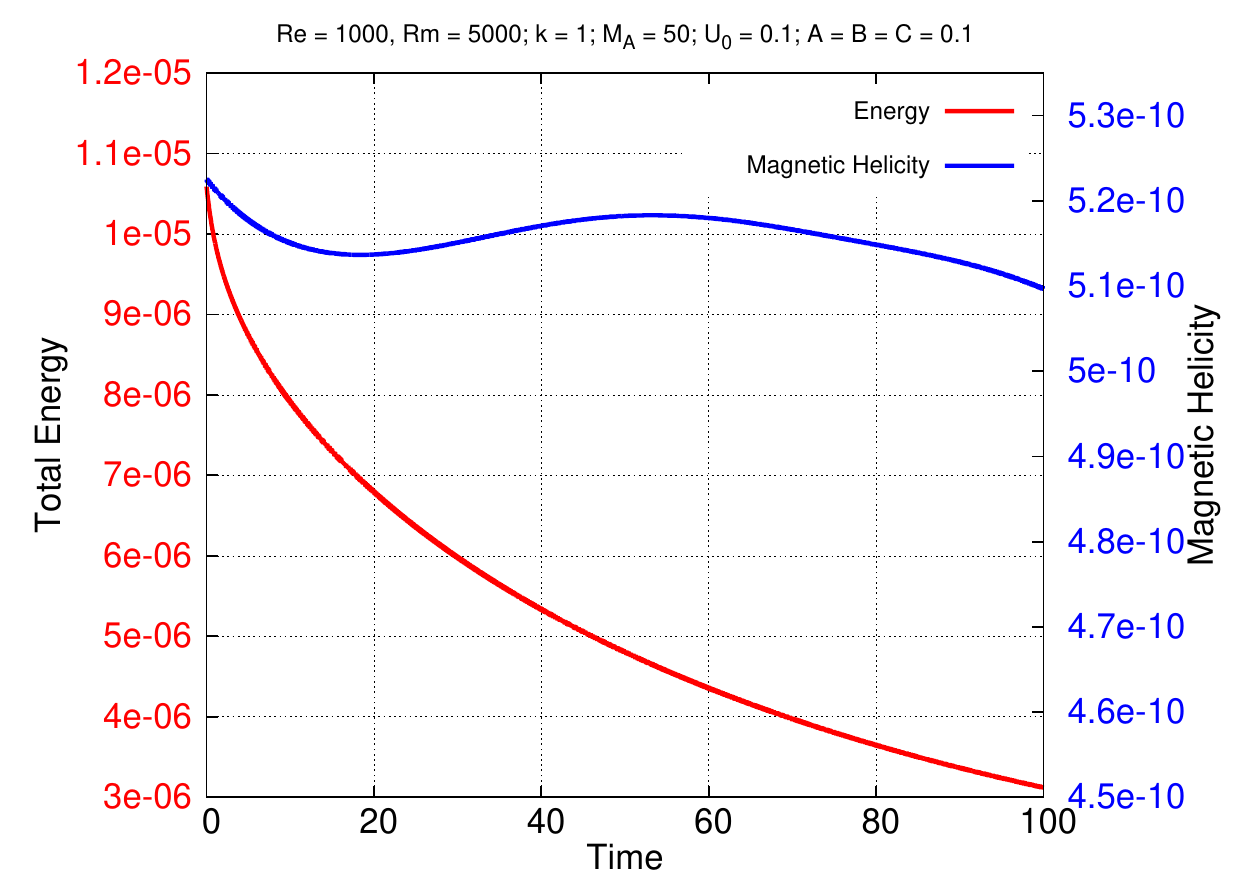}\\
\includegraphics[scale=0.65]{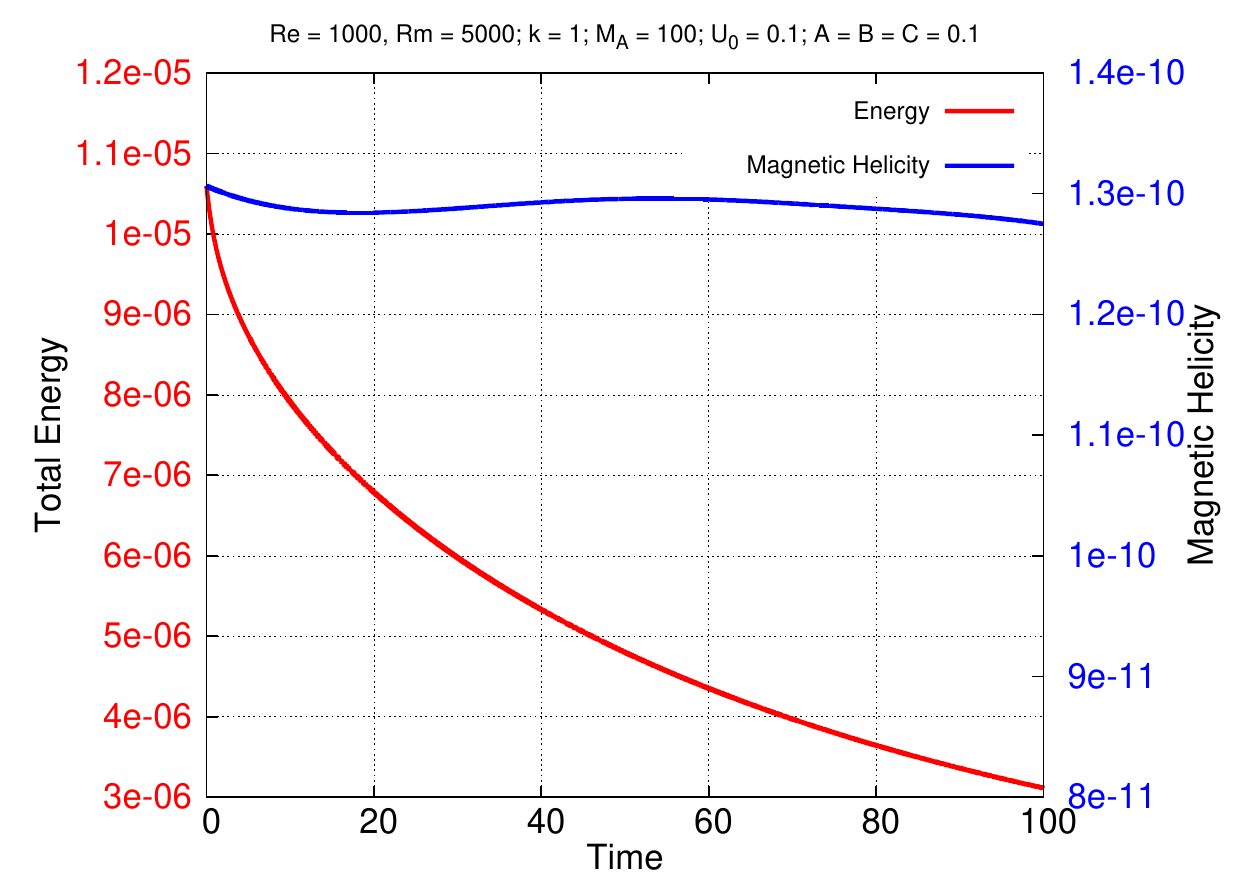}
\includegraphics[scale=0.65]{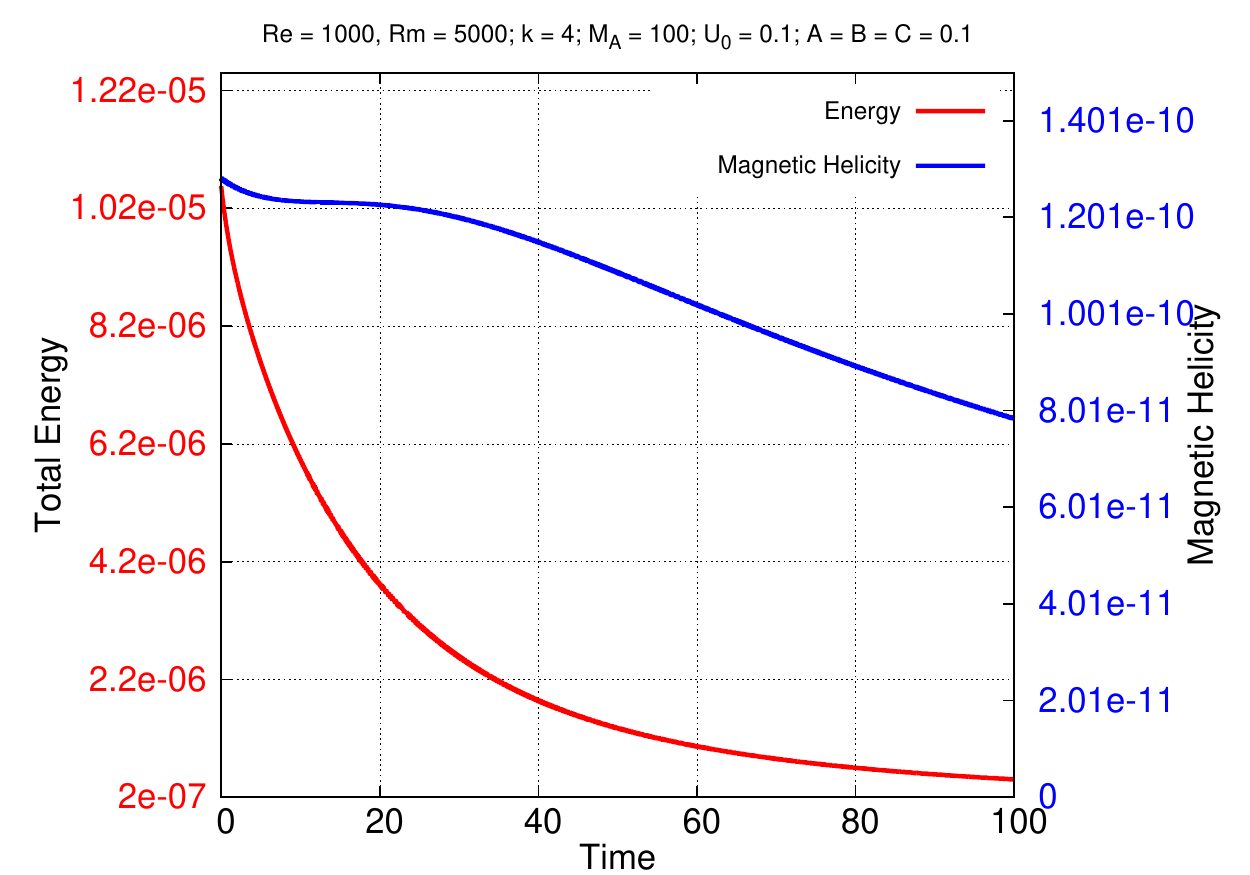}\\
\includegraphics[scale=0.65]{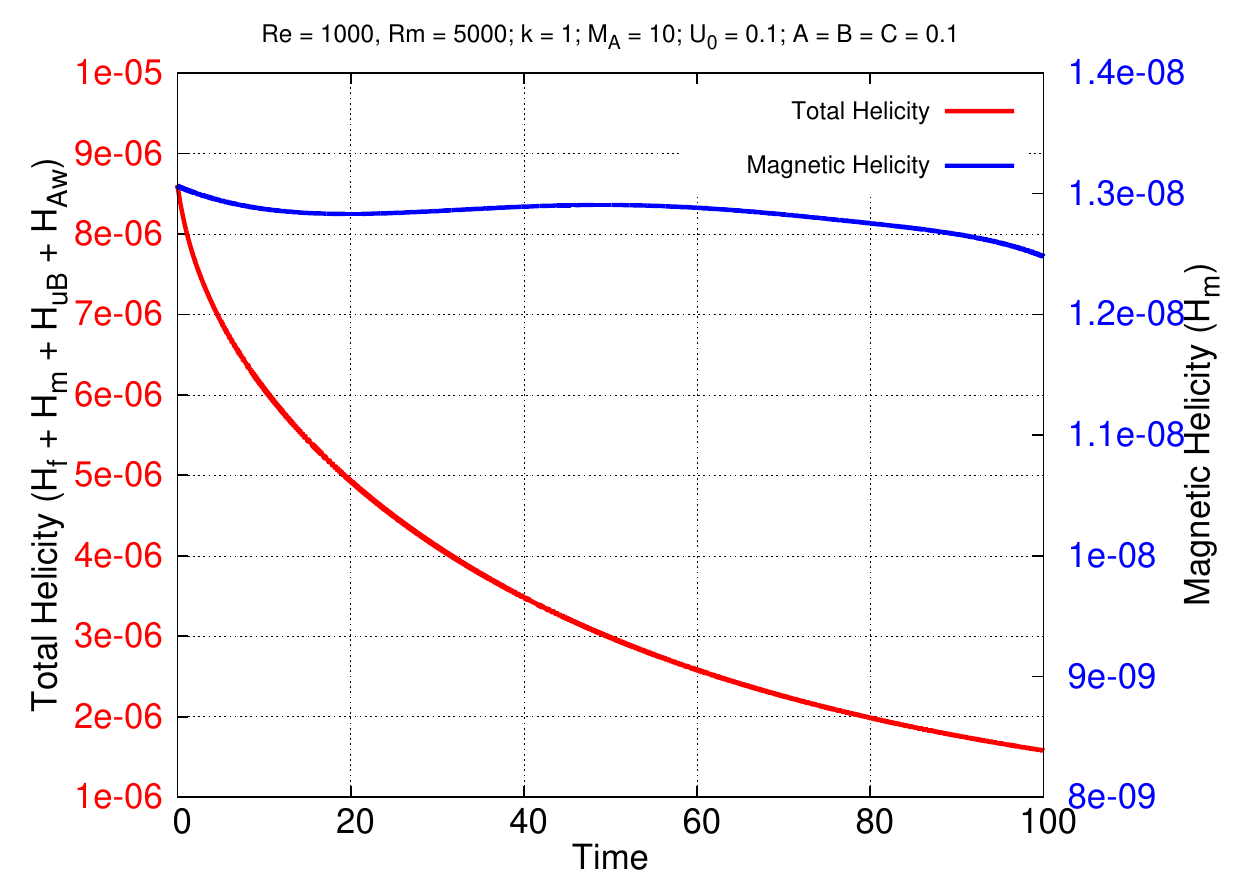}
\includegraphics[scale=0.65]{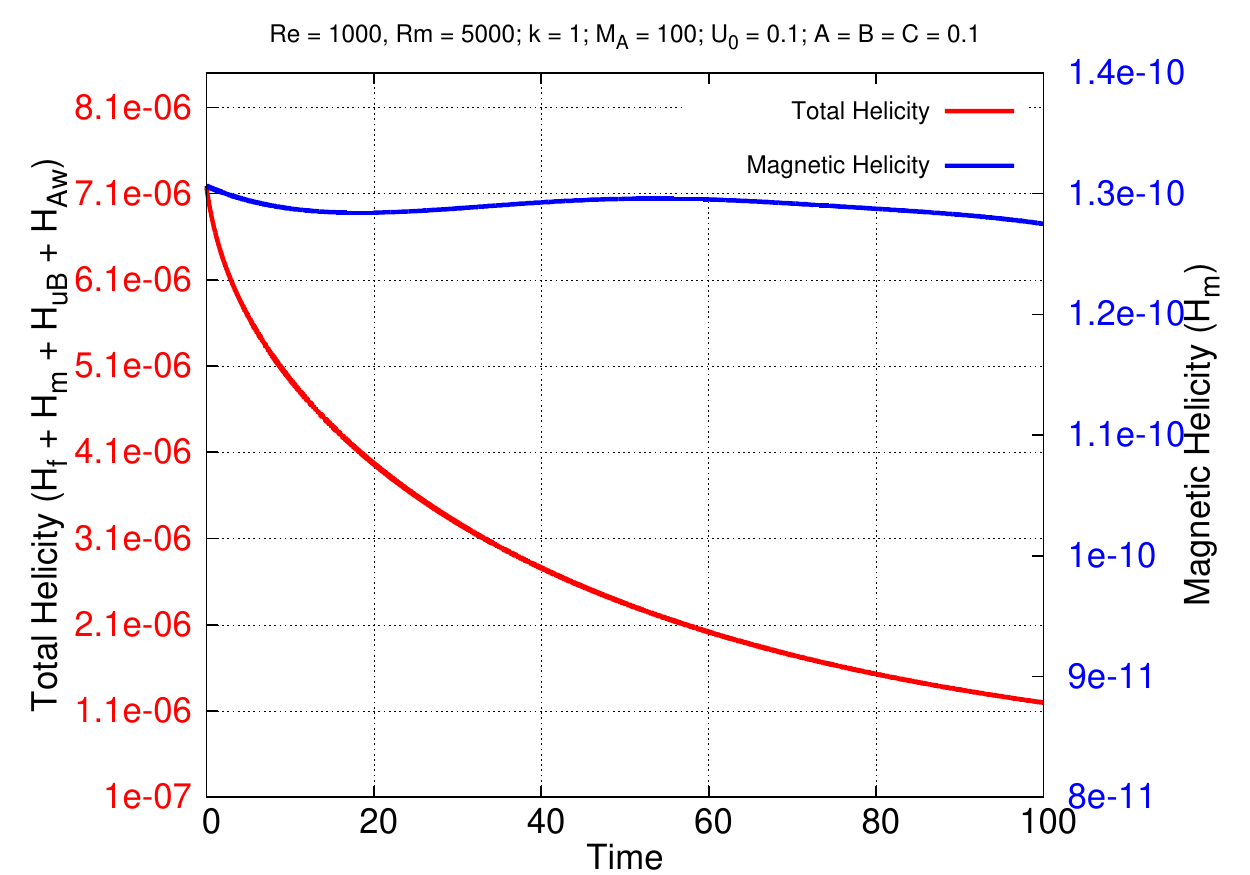}
\caption{Rate of change of Helicity is much smaller compared to rate of change of Energy (Kinetic + Magnetic) and total Helicity ($H_G = H_f + H_m + H_{uB} + H_{A\omega}$) supporting Taylor like situation. The runs are given for the parameters $U_0 = 0.1$, $A = B = C = 0.1$, $Re = 1000$, $Rm = 5000$, $k = 1$ and $M_A = 10, 50, 100$.}
\label{Taylor}
\end{figure}

We run BMHD3D for four values of wavenumber (viz. $k = 1, 2, 4, 8$ at $M_A = 100$) and four values of Alfven Mah number (viz. $M_A = 5, 10, 50, 100$ at $k = 1$) optimised according to the grid resolution we use. Below we provide the time evolution of spatially averaged energy (kinetic + magnetic) $\left( E(t) = \int\limits_V [u^2(t) + B^2(t)] dV \right)$ and magnetic helicity for different values of $M_A$ (Fig.\ref{Taylor}). We observe that the rate of change of magnetic helicity is much smaller compared to rate of change of energy indicating that our parameters are in the Taylor like condition for all the four values of $M_A$. 

\begin{figure}
\includegraphics[scale=0.65]{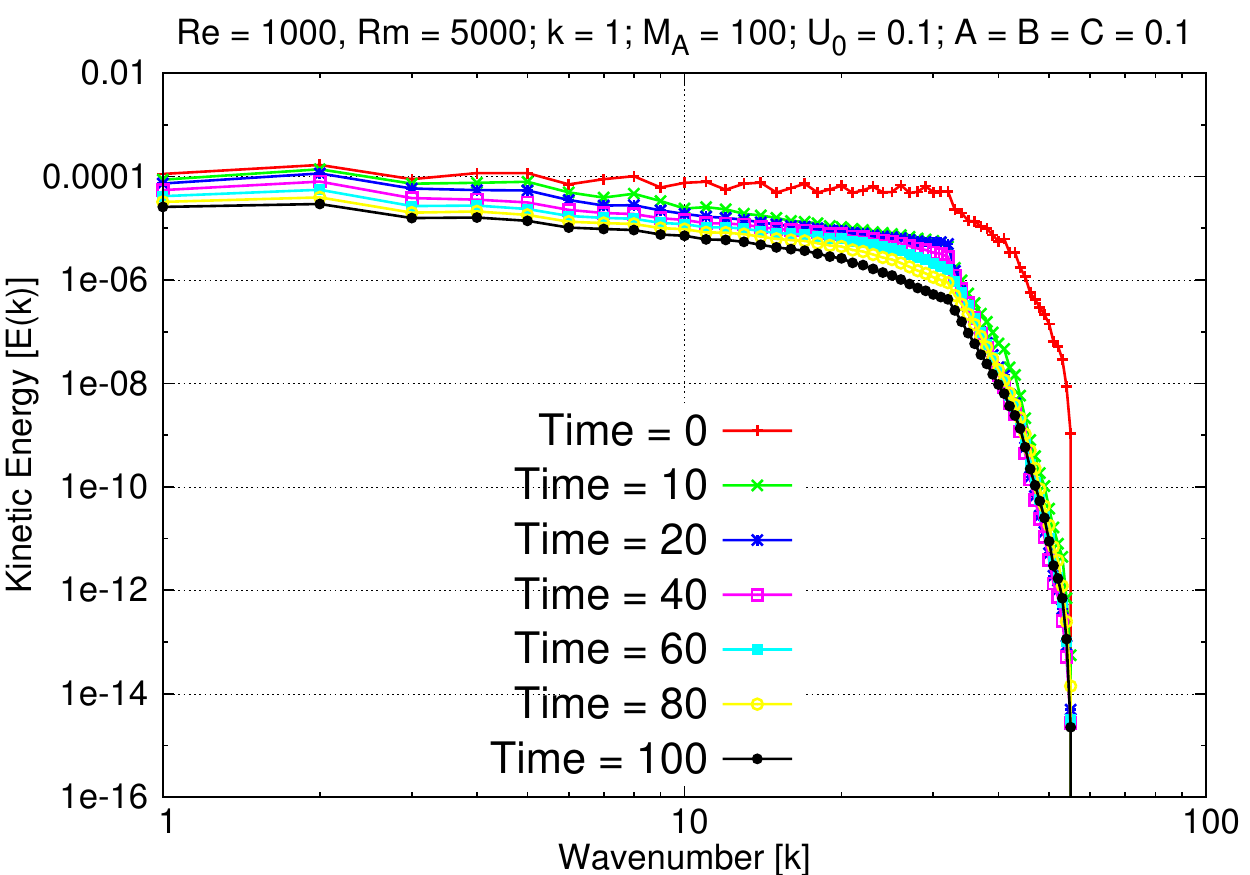}
\includegraphics[scale=0.65]{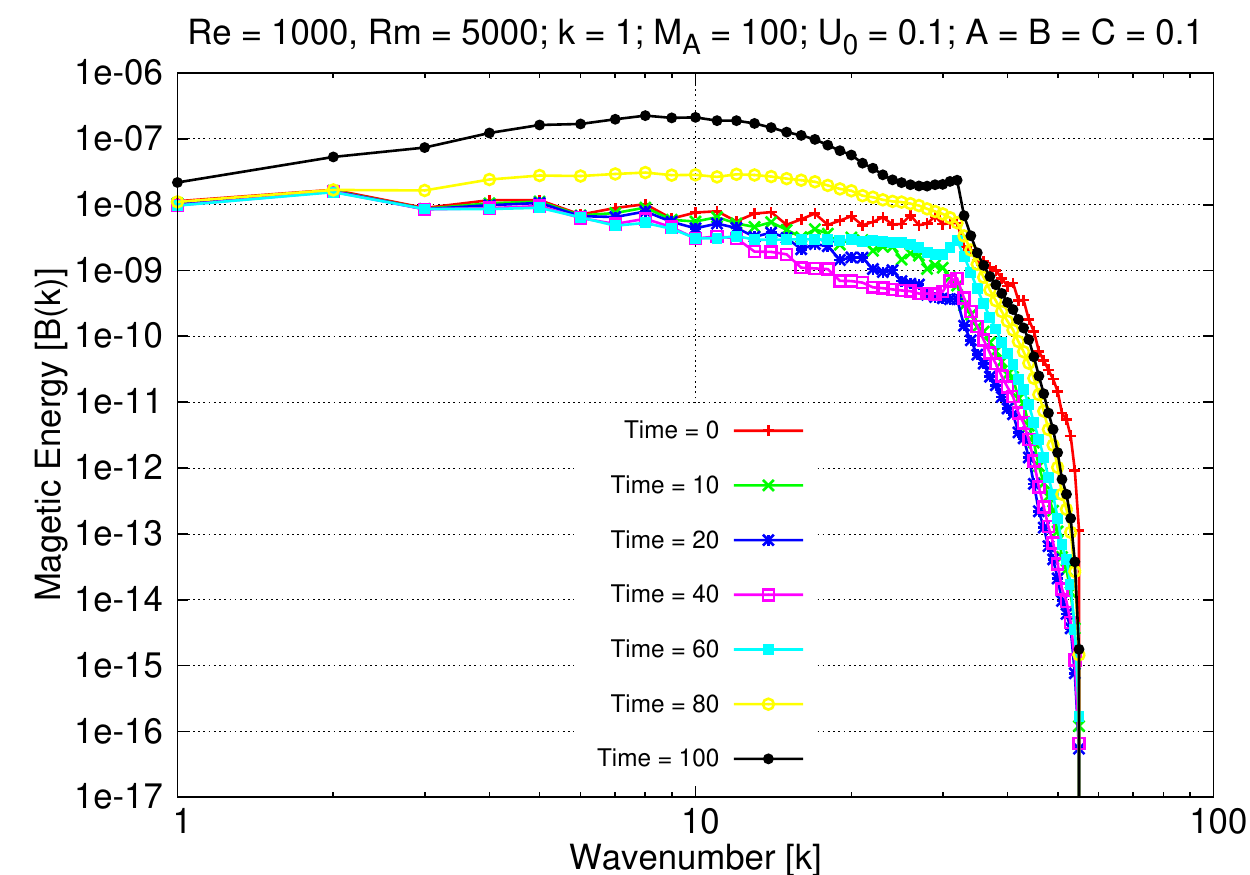}\\
\includegraphics[scale=0.65]{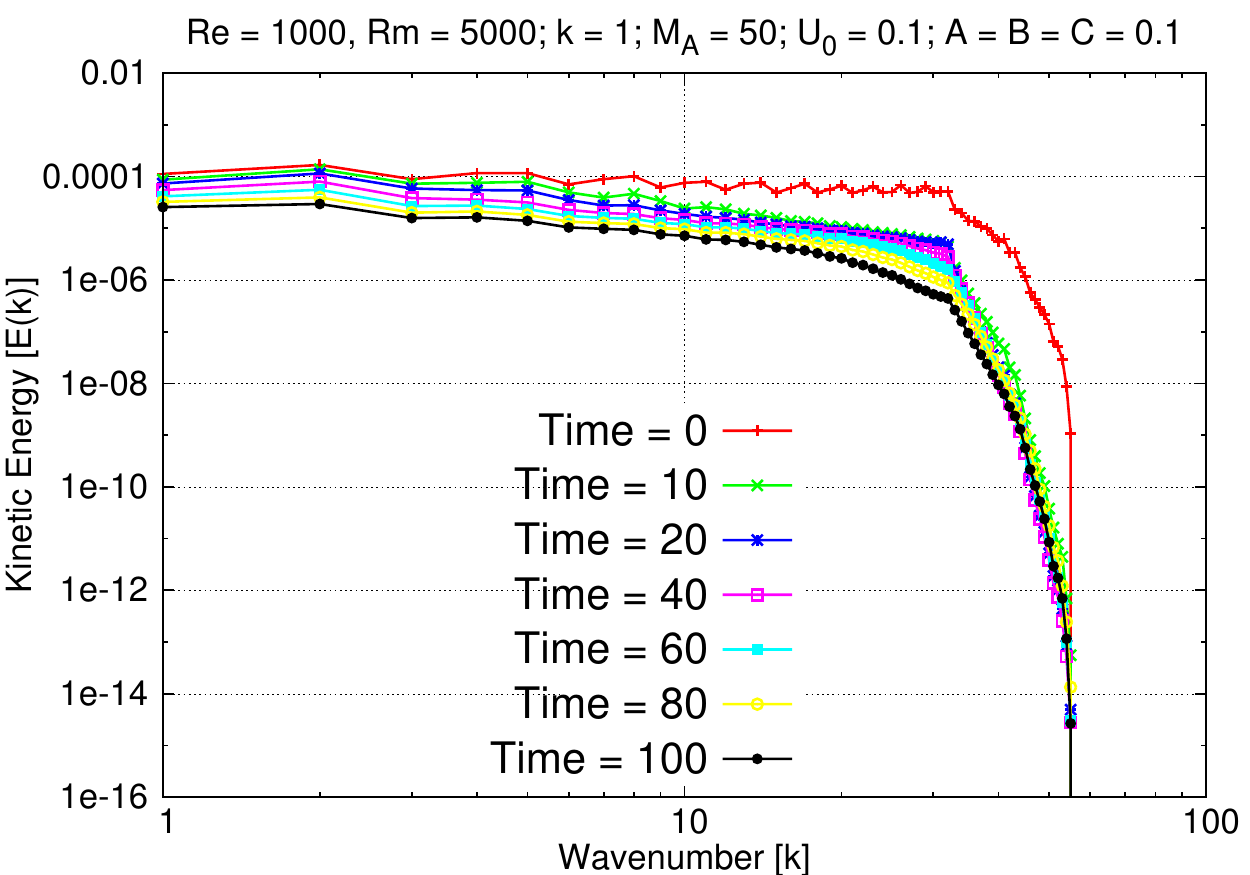}
\includegraphics[scale=0.65]{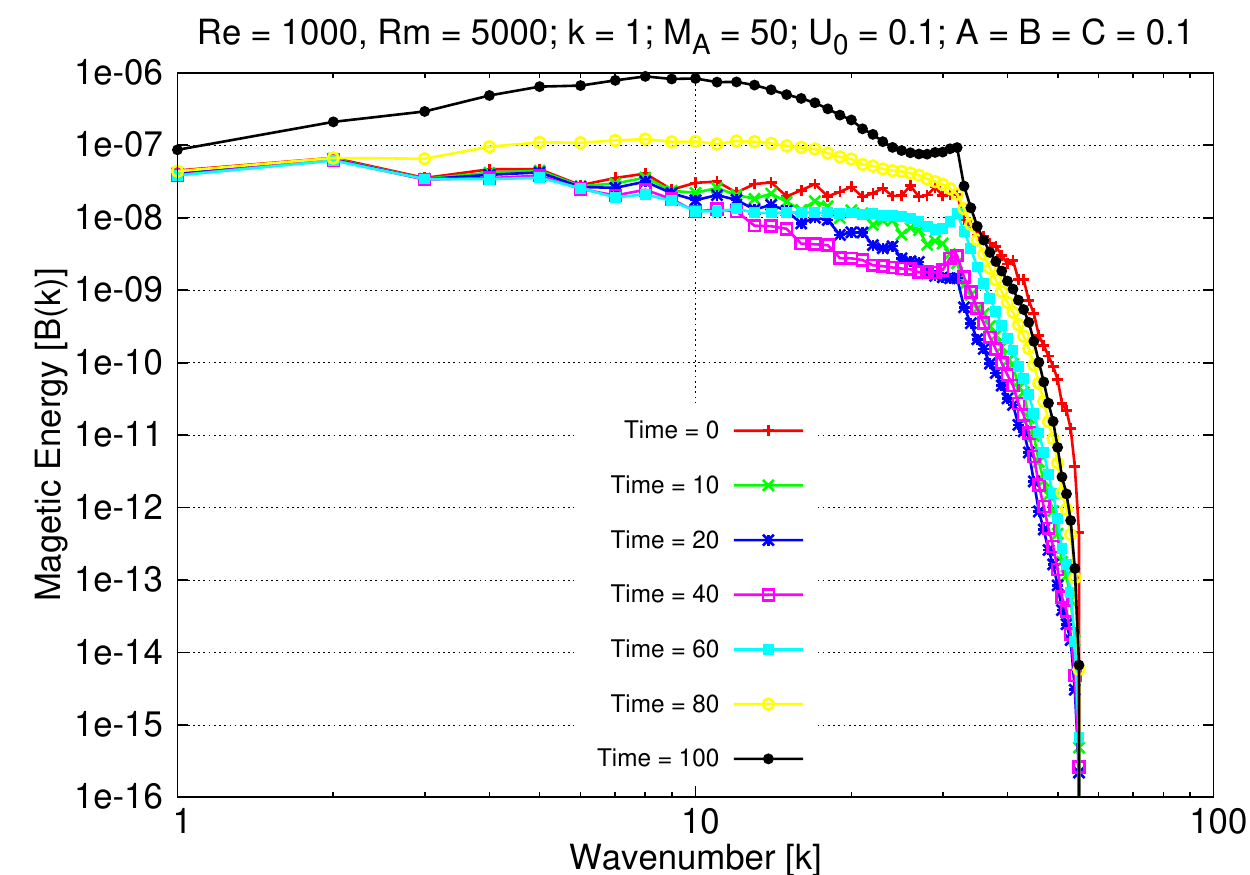}\\
\includegraphics[scale=0.65]{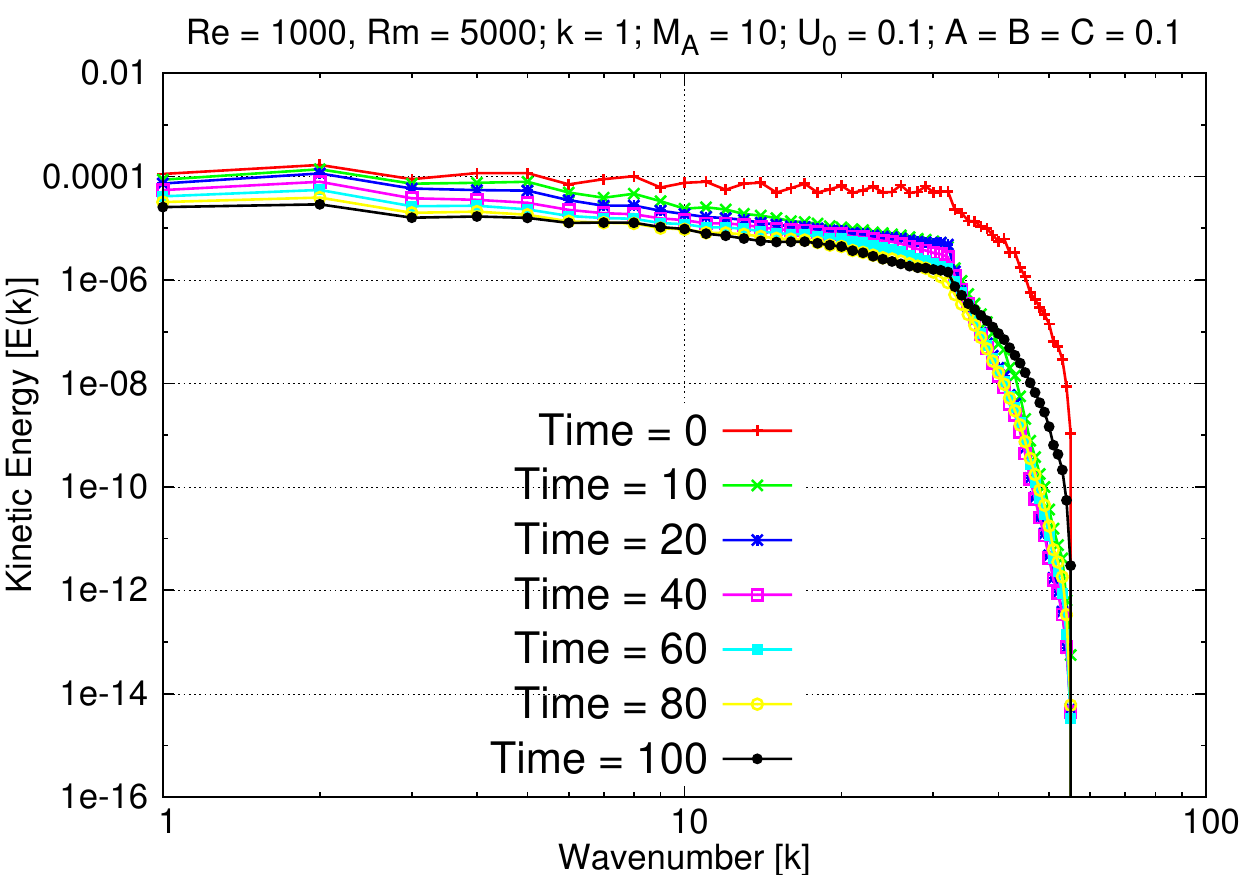}
\includegraphics[scale=0.65]{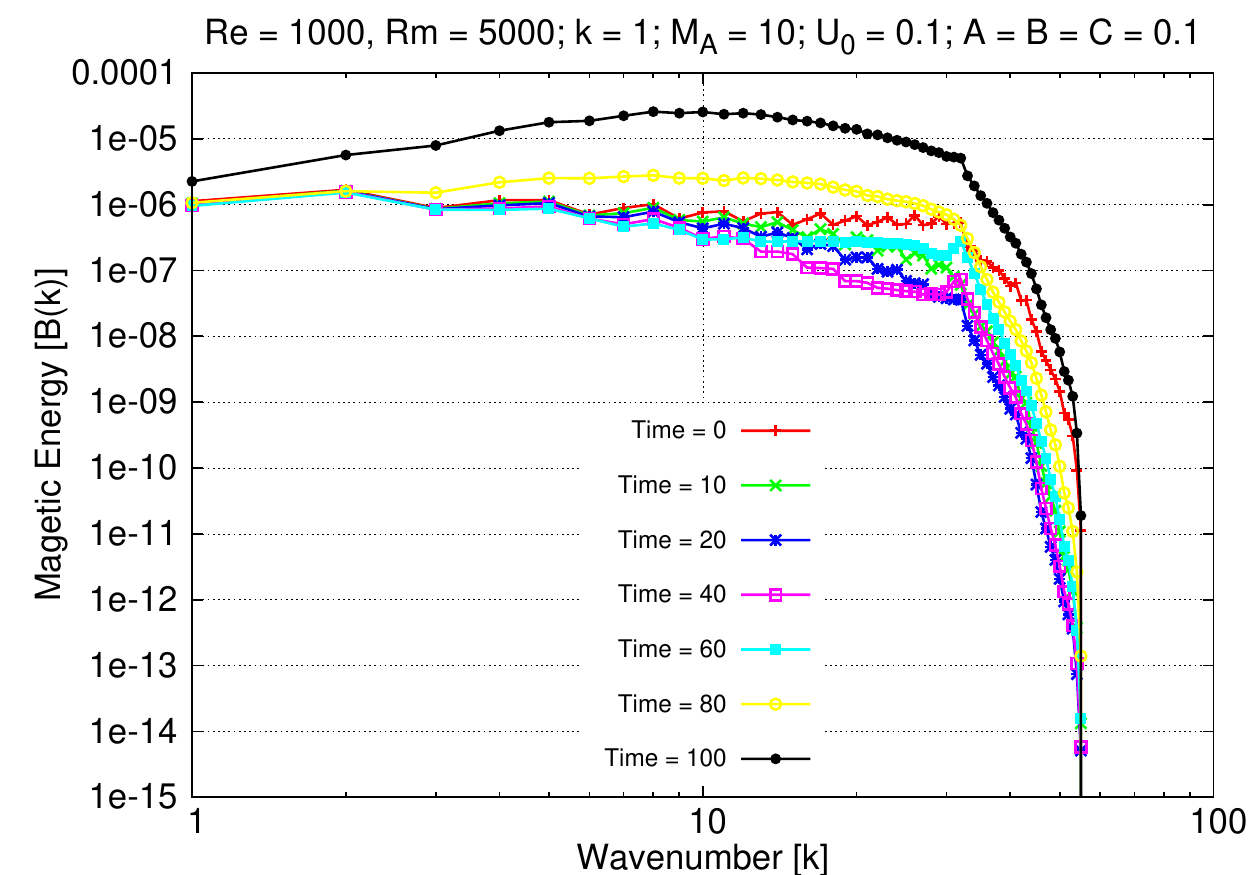}
\caption{The shell averaged kinetic $\left[ \sum u^2 (k_x,k_y,k_z) \right]$ and magnetic $\left[ \sum B^2 (k_x,k_y,k_z) \right]$ energy spectra for different Alfven Mach number ($M_A$). The runs are given for the parameters $U_0 = 0.1$, $A = B = C = 0.1$, $Re = 1000$, $Rm = 5000$, $k = 1$ and $M_A = 100, 50, 10$.}
\label{spectra_MA}
\end{figure}

\begin{figure}
\includegraphics[scale=0.65]{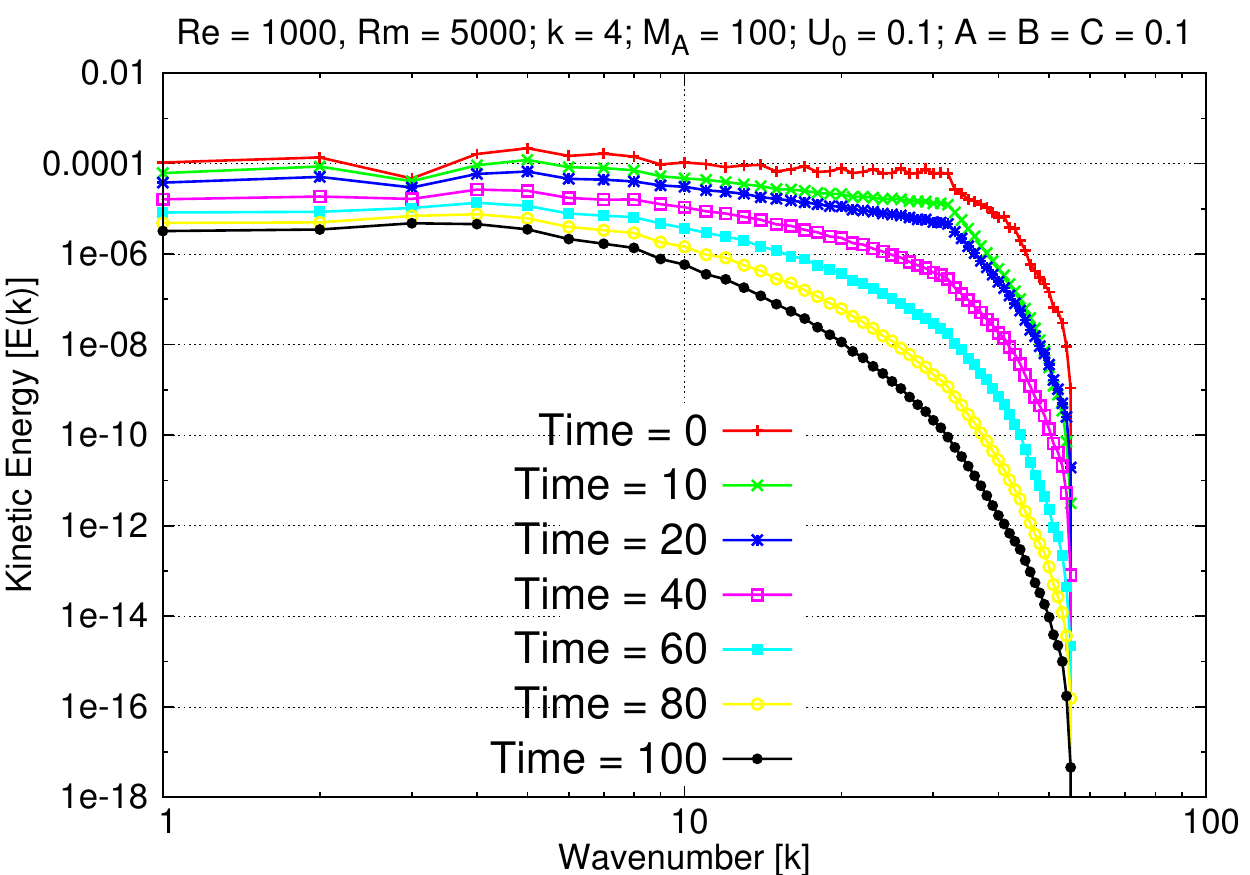}
\includegraphics[scale=0.65]{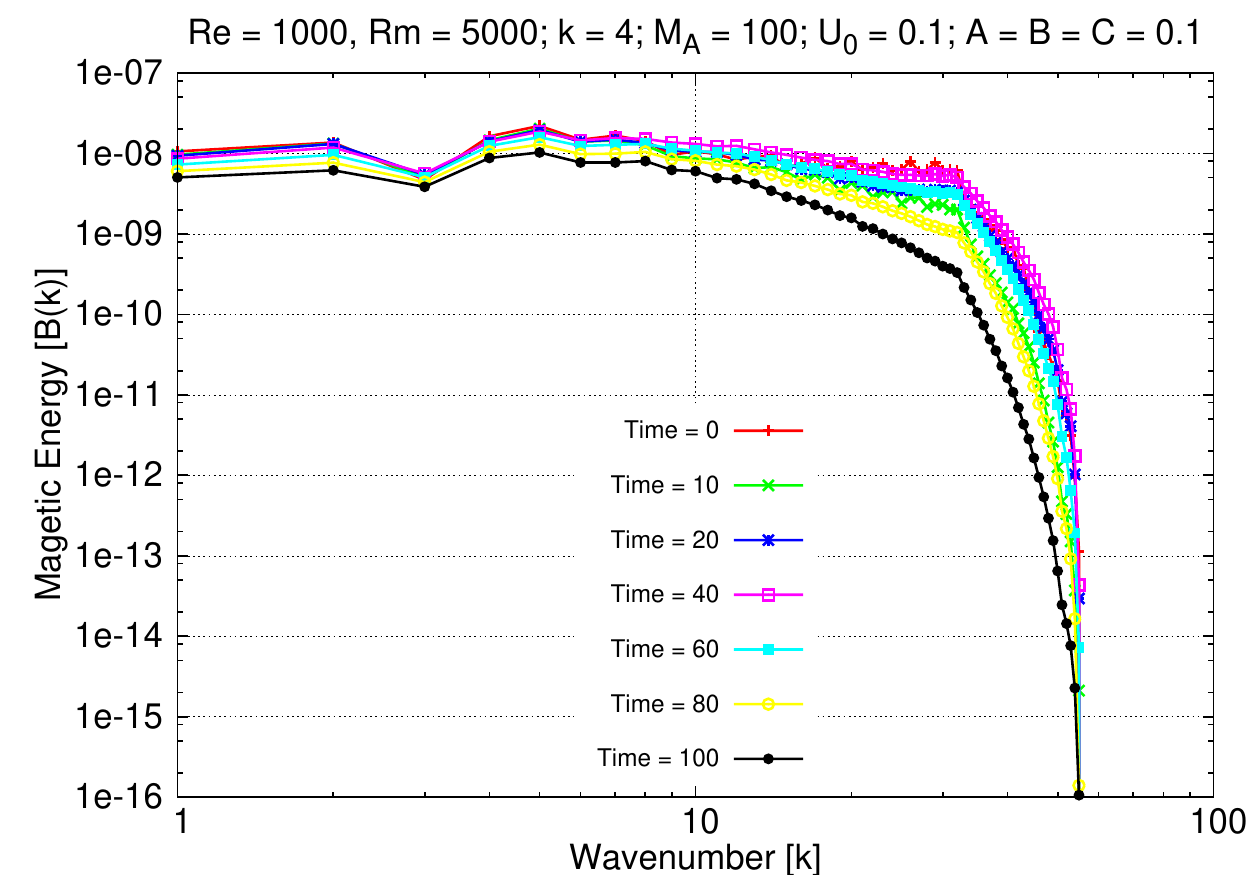}\\
\includegraphics[scale=0.65]{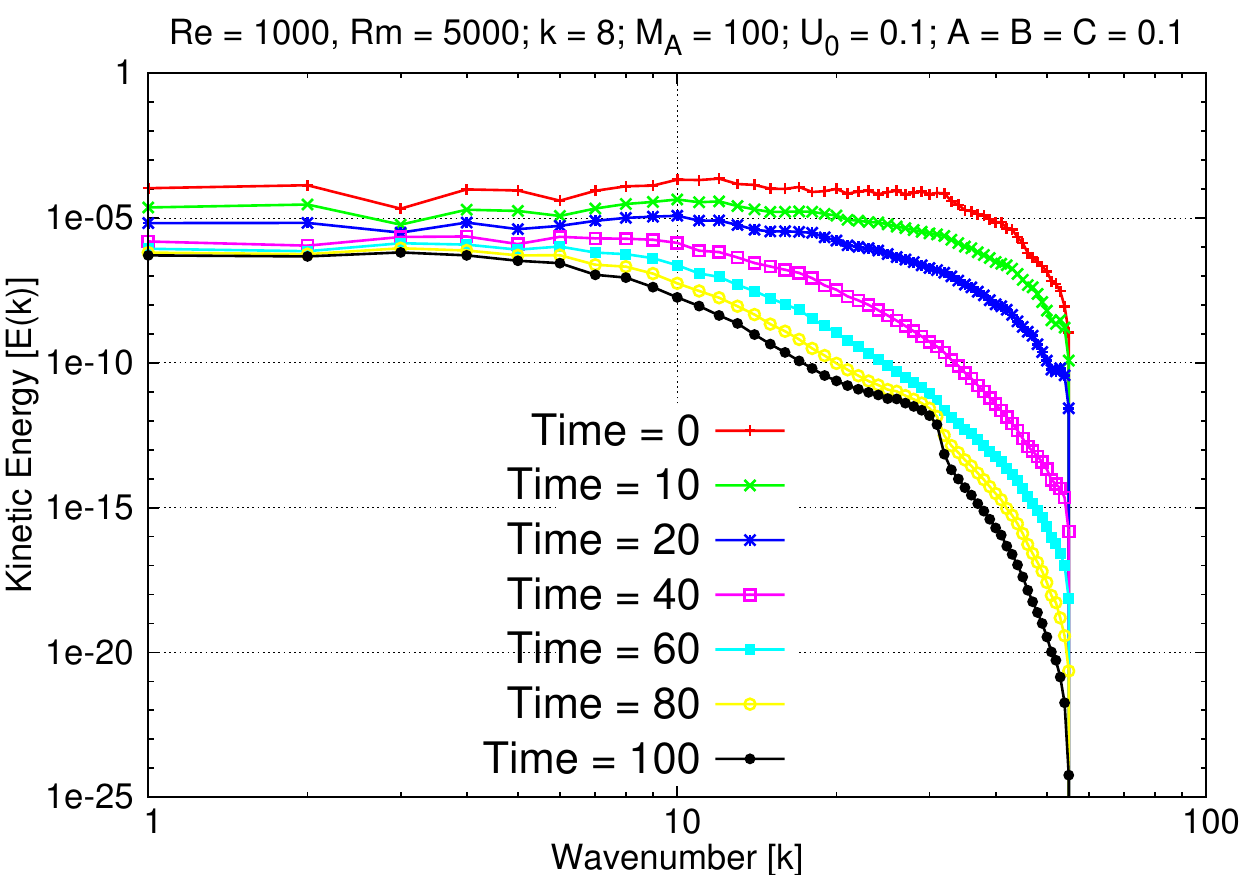}
\includegraphics[scale=0.65]{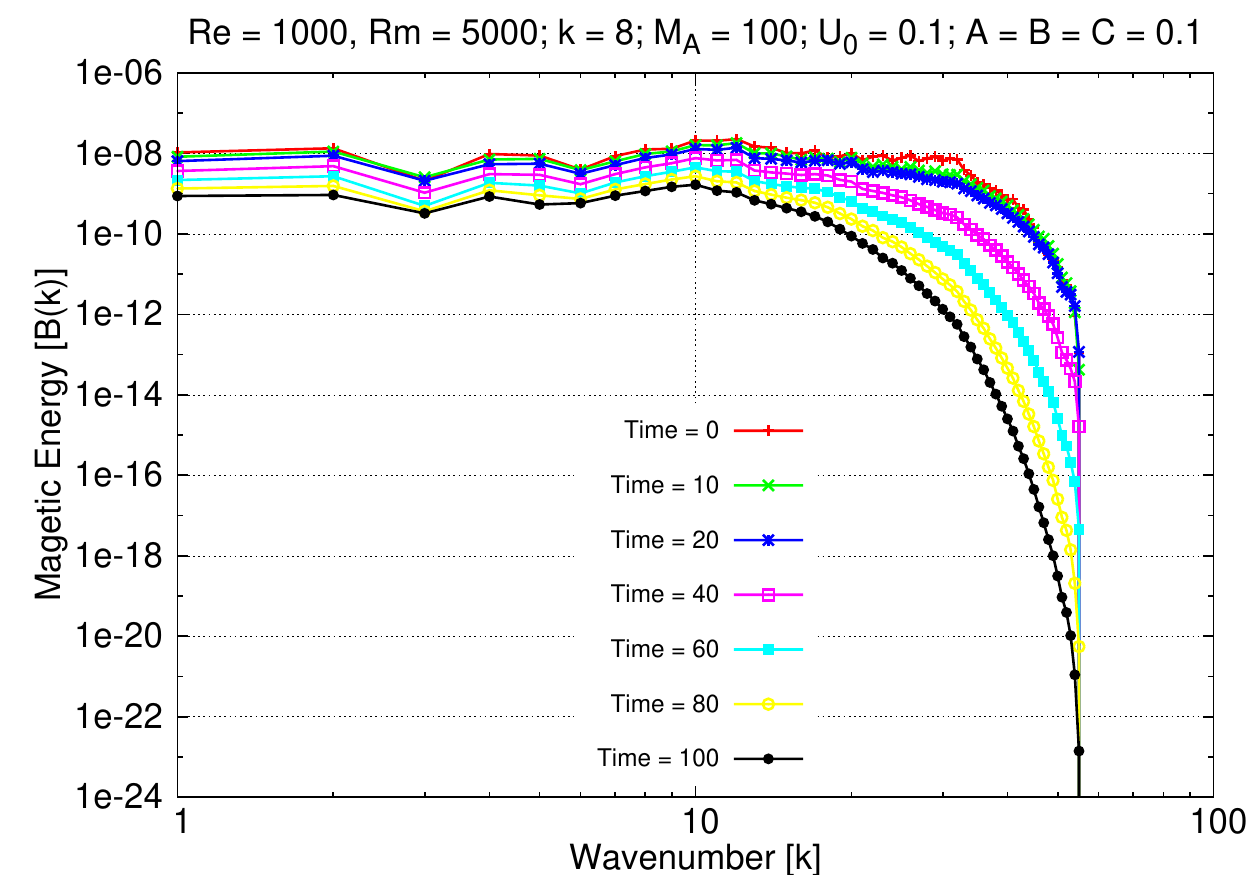}
\caption{The shell averaged kinetic $\left[ \sum u^2 (k_x,k_y,k_z) \right]$ and magnetic $\left[ \sum B^2 (k_x,k_y,k_z) \right]$ energy spectra for different initial wavenumber ($k$). The runs are given for the parameters $U_0 = 0.1$, $A = B = C = 0.1$, $Re = 1000$, $Rm = 5000$, $M_A = 100$ and $k = 4$ \& $8$.}
\label{spectra_k}
\end{figure}

We evaluate the shell averaged kinetic $\left[ \sum u^2 (k_x,k_y,k_z) \right]$ and magnetic $\left[ \sum B^2 (k_x,k_y,k_z) \right]$ energy spectra at different times and observe the time evolution of both of them for different wavenumbers (Fig. \ref{spectra_k}) and Alfven speeds (\ref{spectra_MA}). Though the initial spectra for kinetic and magnetic energy shows identical structure, with the time evolution we notice that they evolve differently.

\section{Summary and Discussion}

In general, as the plasma with flow dynamics included, is allowed to expand suddenly or abruptly from small volume to fill the ideal box ($\vec{B} \cdot \hat{n} = 0 = \vec{u} \cdot \hat{n}$), the expansion process is seen to be strongly dependent on the mode spectrum of the initial magnetic field and velocity field spectra. The results are interesting as they are seen to support a multiscale relaxation process suggested by H. Qin {\it et al} [\ref{Qin:12}].

It is important to note that while our calculation include flow, both Taylor [\ref{Taylor:74}] and H. Qin {\it et al}'s [\ref{Qin:12}] calculations only consider magnetic helicity and energy. It will be interesting to attempt a calculation similar to H. Qin {\it et al} [\ref{Qin:12}] but including flow dynamics. Also, the simulations may yield surprises when the parameters allow total helicity conservation while total energy decays, inline with Mahajan \& Yoshida [\ref{Mahajan:98}] and see if ``Double Beltrami" states appear. Finally, it should be also possible to choose parameters so that, energy dissipation rate is minimised and obtain triple Beltrami states [\ref{Dasgupta:98}].\\

The numerical results we have obtained are for $64^3$ resolution and thus do not capture the dynamics of the energy content at the high wavenumbers. The Galerkin truncation of the discrete sine and cosine transforms used in the code to evaluate the spatial derivatives can have significant effects on the numerical relaxation process we are interested in and can thereby alter the conclusions we have derived from the current results. Upgradation of the code to simulate higher grid sizes are in progress and will be reported elsewhere.

\section{Acknowledgement}

One of the authors RM thanks Udaya Maurya, IPR for his kind help in using FFTW libraries for the discrete sine and cosine transforms. The code BMHD3D has been developed and benchmarked in Uday and Udbhav cluster at IPR.

\end{document}